\documentclass[a4paper,fleqn]{cas-sc}

\usepackage[utf8]{inputenc}

\usepackage[style=numeric, sorting=none]{biblatex}

\bibliography{RTVC}

\usepackage{mathrsfs}

\usepackage{physics}
\usepackage{marginnote}
\usepackage{bbold}
\usepackage{tensor}
\usepackage{caption}
\usepackage{booktabs}
\usepackage[many]{tcolorbox}
\usepackage{empheq}
\usepackage[toc,page, title, titletoc]{appendix}
\usepackage{amsfonts,amssymb,amsmath,amsthm,amstext,amssymb,amsopn,mathtools,nicefrac,xfrac}
\usepackage{bm}

\usepackage{slashed}
\usepackage{subcaption}
\usepackage{float}
\usepackage{hyperref}
\usepackage{tcolorbox}
\usepackage[]{todonotes}
\usepackage{graphicx}
\usepackage{mathtools} 
\usepackage{cleveref}
\usepackage{url}
\usepackage{algorithm}
\usepackage{algpseudocode}
\usepackage{listings}

\begin{document}
\let\WriteBookmarks\relax
\def\floatpagepagefraction{1}
\def\textpagefraction{.001}
\shorttitle{Real-time Plasma Shape Control in PCS}
\shortauthors{M. Marshall, E. Jones, et~al.}

\title[mode=title]{Real-time virtual circuits for plasma shape control via neural network emulators: integration and testing in the MAST-U PCS}

\affiliation[1]{
    organization={Hartree Centre, STFC},
    addressline={Daresbury Laboratory},
    city={Daresbury},
    postcode={WA4 4AD},
    country={United Kingdom}
}

\affiliation[2]{
    organization={UKAEA},
    addressline={Culham Campus},
    city={Abingdon},
    postcode={OX14 3DB},
    country={United Kingdom}
}

\author[1]{Matthew J. Marshall}[orcid=0000-0003-1454-2890]
\ead{matthew.marshall@stfc.ac.uk}

\author[2]{Edward Jones}
\ead{edward.jones1@ukaea.uk}

\author[2]{Graham J. McArdle}

\author[1]{Alasdair Ross}

\author[2]{Kamran Pentland}

\author[2]{Nicola C. Amorisco}

\author[2]{Charles Vincent}

\author[2]{Martin Kochan}

\author[2]{Colin Hogben}

\author[2]{Graham Jones}

\author[2]{Adam Stephen}

\author[2]{George K. Holt}[orcid=0000-0001-6814-9117]

\author[1]{Adriano Agnello}[orcid=0000-0001-9775-0331]


\begin{abstract}
The deployment of advanced, AI-enabled control algorithms in tokamak experiments requires robust integration with existing plasma control system (PCS) architectures and extensive pre-experimental validation.
In this contribution, we describe the integration and testing of neural-network-emulated virtual circuits for plasma shape control within the MAST Upgrade (MAST-U) PCS environment.
The neural network models predict the plasma shape using the plasma current, poloidal field coil currents, and plasma profile parameters.
In this paper, we explain how they are deployed via a real-time C++ inference server that interfaces with the PCS, returning the shape prediction and its Jacobian, and how, from the latter, virtual circuit matrices and updated coil current requests are computed for real-time actuation.
Emphasis is placed on the validation workflow and best practices adopted to ensure confidence in the proposed control framework prior to experimental deployment.
This work demonstrates practical AI-based shape control components for fusion control systems, with direct relevance for upcoming MAST-U experiments and future devices.
\end{abstract}

\maketitle

\section{Introduction}
\label{s-intro}

To achieve net energy production, future magnetic fusion devices will require accurate and responsive plasma position and shape control in order to access high performance operational regimes \cite{sips_advanced_2005,meyer_physics_2022}.
Recent work has shown that machine-learning methods can be integrated into plasma control systems (PCS) at several different levels of the control stack. 
Early neural-network (NN) work already established the feasibility of real-time learned feedback for tokamak position and shape control \cite{bishop1995realtime}. 
More recently, reinforcement-learning controllers have been trained in simulation and deployed directly on TCV for magnetic control of diverse plasma configurations, including advanced shapes \cite{degrave2022magnetic}, with subsequent work addressing practical issues such as training time, steady-state offsets, and routine controller generation \cite{tracey2024practical}. 
Related developments on DIII-D have demonstrated reconstruction-free magnetic control using raw magnetic diagnostics in the PCS \cite{subbotin2026diii}, while broader AI-control infrastructure has enabled the deployment of multiple machine-learning predictors and controllers within integrated real-time experiments \cite{rothstein2026pacman}. 
These results highlight that the central challenge is not only the learning algorithm itself, but the full deployment chain: simulator-based testing, real-time-compatible execution, diagnostic and actuator interfacing, constraint handling, and experimental validation. 

In this paper, we detail the integration and testing of a new AI-enabled plasma shape controller within the MAST Upgrade (MAST-U) PCS. 
The NN-based shape controller takes in real-time measurements of the plasma current, poloidal field (PF) coil currents, and plasma current density profile parameters, returning state-aware virtual circuits (VCs) to the PCS, which are used to compute updated PF coil current requests for real-time shape actuation. 
Compared to the conventional VCs, computed offline prior to a plasma shot, these emulated VCs provide state-aware shape sensitivity, enabling adaptivity to unforeseen plasma trajectories on milliseconds timescales---a capability well-suited to MAST-U which explores a wide range of highly shaped plasma configurations \cite{harrison_overview_2024,anand_real-time_2024}. 

This work completes the validation process for AI-enabled shape control, building on the prior validation of the emulated VCs accuracy against Grad-Shfranov (GS) solutions in static scenarios \cite{ross_real-time_2026} and in closed-loop simulations \cite{pentland_real-time_2026}.
Here, we demonstrate that the fully PCS-software integrated real-time VC (RTVC) controllers produce the expected output during pre-shot testing and ``piggyback'' commissioning shots (to $R^2 > 0.99$ levels), while retaining the conventional shape control capabilities. 

In Section \ref{s-existing-shape-control}, we review the existing shape control capability on MAST-U PCS.
In Section \ref{s-rtvc-server}, we detail the architecture of the RTVC Server, the system that serves real-time inference of NN models, covering the real-time constraints, the communication model between it and PCS, and the outputs produced. 
In Section \ref{s-integration-and-testing}, we detail the RTVC Algorithm that integrates RTVC Server into PCS, what transformations are made for downstream control, and in-silico testing of this integration with timing and command validation. 
In Section \ref{s-irl-testing}, we present final validation of the control scheme in several commissioning shots on MAST-U.
Finally, in Section \ref{s-conclusion}, we discuss opportunities there are to improve on the control scheme, including generalisation for RTVC Server's use to other aspects of tokamak control.

\section{Plasma Shape Control in the MAST-U PCS}
\label{s-existing-shape-control}

The MAST-U PCS is built on the General Atomics real-time control framework \cite{penaflor_worldwide_2008}, which is also used on DIII-D, NSTX-U, EAST, and KSTAR \cite{mcardle_mast_2020}.
Real-time plasma shape reconstruction is performed by LEMUR (Local Expansion for MAST Upgrade Reconstruction) \cite{kochan_real-time_2023}, which uses a local flux-expansion method with magnetic measurements to locate a set of plasma shape descriptors. 
These descriptors, collected in vector $\vec{P}$, include the inboard and outboard midplane radii ($R_{\textbf{in}}$, $R_{\textbf{out}}$), X-point position ($R_X$, $Z_X$), divertor nose radius ($R_{\textbf{nose}}$), strike-point radius ($R_{\textbf{strike}}$), and squareness gap ($S_{\textbf{gap}})$ \cite{mcardle_integrated_2023}.

A VC matrix is defined as the pseudoinverse of the sensitivity matrix $S = \partial \vec{P}/\partial \vec{I}_\mathrm{shape}$, where $\vec{I}_\mathrm{shape}$ is the vector of PF coil currents.
It maps requested changes in each shape parameter to the corresponding combination of coil current changes needed to produce it, leaving the other shape parameters unchanged.
Requested changes $\partial \vec{P} / \partial t$ from the shape controller are multiplied by a VC at each time step to obtain requested changes in coil currents $\partial \vec{I}_\mathrm{shape} / \partial t$. 
Machine-protection limits are applied to these requests before they are converted to PF coil voltages.

VCs are typically computed offline, via finite-differences, using a small number of GS equilibria taken along the anticipated scenario trajectory, with each VC applied over a preset time interval. This complicates planning scenarios in which the plasma shapes evolves substantially. Furthermore, in case of departures from the anticipated trajectory, the accuracy of the VC degrades and control performance suffers. 
Since computing $S$ using GS solutions and inverting it is too slow for real-time deployment, we use the trained emulators from \cite{ross_real-time_2026} to predict $\vec{S}$ and generate the VCs in real-time. 
These require real-time measurements of $\vec{I}_\mathrm{shape}$, the total plasma current, and the plasma current density profile parameters.

\section{Real-time Virtual Circuit Server}
\label{s-rtvc-server}

Any control scheme for the MAST-U PCS must meet basic real-time constraints: operating system (OS) level memory allocation confined to initialisation, low-latency inter-process communication that avoids OS-level routine calls (syscalls), and a sufficiently fast cycle time. 
The aforementioned NN emulators were trained using TensorFlow \cite{abadi_tensorflow_2016} and Keras \cite{chollet_keras_2018}, and were validated using their native Python/Keras interface \cite{ross_real-time_2026,pentland_real-time_2026}.
This interface is unsuitable for real-time use: inference times are long and jittery, and memory allocation cannot be constrained.

We identified TensorFlow Lite for Microcontrollers\footnote{Now known as LiteRT for Microcontrollers \cite{wang_litert_2026}.} (TFLM) \cite{david_tensorflow_2021} as a suitable inference engine, since it targets microcontroller environments with constraints similar to PCS (differing mainly in available memory, processor power, and PCS's x86-64 architecture).
Alternatives such as ONNX \cite{noauthor_onnx_nodate}, OpenVINO \cite{zunin_intel_2021}, and NCNN \cite{ni_ncnn_2017} were ruled out as they could not avoid memory allocation during inference, while TVM AOT \cite{chen_tvm_2018} had steeper model development and integration times.
After NN training, Keras models are easily converted to TFLM's flatbuffer fomrat using its tooling; these flatbuffers, plus a short descriptor file, are fed to RTVC Server and stored in a dedicated mount on PCS hardware.

\begin{figure}
    \centering
    \includegraphics[width=0.98\linewidth]{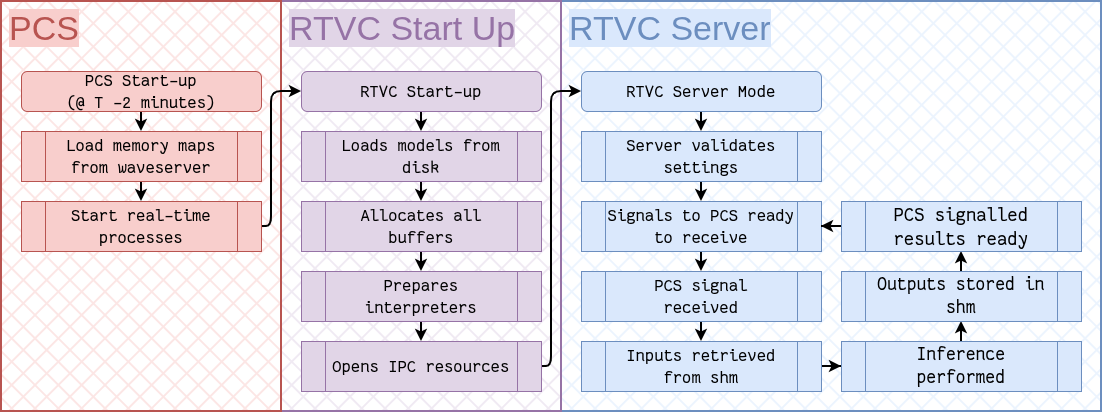}
    \caption{RTVC Server architecture, showing related PCS initialisation (red), RTVC Server initialisation (purple), RTVC Server "server mode" (blue). All memory allocations occur in the red and purple phases, with the blue phase simply performing inferences and finite-differences, retrieving inputs from and sending results to a shared memory segment.}
    \label{fig:rtvc-server-architecture}
\end{figure}

Figure \ref{fig:rtvc-server-architecture} shows RTVC Server's two components: command-line interface (CLI) program (purple) and library (blue), as well as the associated PCS initialisation system (red).
The CLI validates inputs and allocates memory, then hands off to a library-provided mode, after which NN inference begins.
RTVC Server runs in three modes: shape-parameters only, Jacobian, and multi-threaded Jacobian.
The first two are single-threaded; shape-parameters-only returns just the emulator-predicted shape parameters, while the other two also return the emulators' Jacobian. TFLM (like all real-time-suitable inference engines considered here) lacks autodiff support, but tuned finite differences were found to give comparable or better accuracy to autodiff Jacobians \cite{ross_real-time_2026}, making this our preferred approach for this implementation.

RTVC Server can also load an ensemble of models, averaging their shape parameters and Jacobian predictions to trade accuracy against extra computational time (i.e. the achievable VC update rate).
In multi-threaded mode, the perturbed inferences needed for finite-difference gradients are distributed across worker threads, while the main thread computes each model's baseline inference and collects the results to form the averaged Jacobian.
With $N$ models, the fastest net inference time scales linearly with the sequential baseline inference.

% timing tests
Timing tests (see Table \ref{tab:rtvc-server-timing}) were run on a development machine with similar hardware to PCS covering three scenarios: no inference, single model, and ensemble of $8$.
Each test spans the full round trip from sending inputs to RTVC Server to retrieving outputs.
With no inference, RTVC Server achieves sub-microsecond latency except at the 99th percentile, averaging 1.5MHz---confirming interprocess communications are sufficient for PCS's 1MHz control cycle.
With a single model, RTVC Server exceeds PCS's 10kHz shape-control frequency up to the 90th percentile and stays sub-$0.2$ms at the 99th. 
With the 8 model ensemble (all results hereafter are generated using this), timings remain sub-$1$ms at the 99th percentile. 

PCS and RTVC Server communicate via two shared memory segments, one for data and one for synchronisation flags (using flags instead of OS semaphores avoids syscalls, cutting latency and jitter), with dedicated isolated cores letting OS threads wait rather than sleep on synchronisation. 
RTVC Server allocates a fixed 1GB model buffer plus 100kB per model for intermediate buffers, so our 8-model ensemble uses under 1.1GB.

\begin{table}[b!]
    \centering

    \begin{tabular}{l|cc|cc|cc|cc}
        \multirow{2}{*}{\textbf{Description}} & \multicolumn{2}{c|}{\textbf{Average}} & \multicolumn{2}{c|}{\textbf{p90}} & \multicolumn{2}{c|}{\textbf{p95}} & \multicolumn{2}{c}{\textbf{p99}} \\
        & ns & kHz & ns & kHz & ns & kHz & ns & kHz \\ \hline
        No Inference & 650 & 1500 & 830 & 1200 & 920 & 1100 & 1200 & 833 \\
        Single Model & 50000 & 20 & 65000 & 15 & 65000 & 15 & 190000 & 5 \\
        Ensemble of 8 & 590000 & 1.7 & 730000 & 1.4 & 890000 & 1.1 & 930000 & 1.1 \\
    \end{tabular}
    \caption{Timings and frequencies of RTVC Server for end-to-end requests (average and 90th, 95th, 99th percentiles).}
    \label{tab:rtvc-server-timing}
\end{table}

\section{PCS Integration and Testing}
\label{s-integration-and-testing}

Control functionality in the MAST-U PCS is organised into independently-scheduled ``Categories'' (e.g. reconstruction, shape control, divertor control), each of which steps through a sequence of ``Phases'' during a pulse, with configuration data and an associated ``Algorithm'' attached to the Category at each Phase \cite{mcardle_mast_2020,mcardle_integrated_2023}.
Conventional shape control uses the VC Algorithm, selecting VCs in real time from a pre-set schedule, and runs directly on CPU2 of PCS at a 0.1ms (10kHZ) cycle time. 

To enable the NN emulated VCs, we developed the RTVC Algorithm (and Category) for selection during a given shot Phase. 
It sends the required real-time measurement data to RTVC server for inference, which returns the ensemble-averaged Jacobian $S$, the pseudoinverse of which (i.e. the VC) is then calculated using CBLAS.
Entries in the VC Algorithm's VC are then overridden with the corresponding entries from this real-time VC for the selected shape parameters (subsets of which can be chosen). 
This overwriting requires the VC Algorithm to be running in the background of the RTVC algorithm when enabled. 

This computational work is split across two PCS cores. 
CPU2 (0.1ms / 10kHz) handles the VC entry overriding and PF coil current request calculation, identical to the VC Algorithm, while CPU5 (0.5ms / 2kHz) handles the RTVC Server signalling, sending inputs/outputs, and performing the pseudoinverse calculation. 
As mentioned above, RTVC Server itself runs on separate pinned threads on cores unused by PCS, asynchronously rather than on a fixed cycle, yielding fast inference times. 
We note that while the first pseudoinverse call incurs a high execution time ($\sim 80~\mu$s allocating worker threads), this happens only once, with the subsequent calls occurring in $8$-$12~\mu$s. 

Build infrastructure to target PCS hardware optimisations, such as vectorisation, are planned for future work. In this implementation, generic Linux binaries were produced and used instead. 
In the future, architecture-optimised binaries will be generated and benchmarked---we expect the timings in Section \ref{s-rtvc-server} to be achievable on PCS hardware, with further reduced jitter. 

Following successful unit testing, we built a dedicated test suite for control-level verification of the complete PCS shape control algorithm, mocking the PCS real-time heap and associated data structures to emulate shot execution and support debugging. Using this data, MAST-U shots can be ``re-played'' on development machines (equivalent to PCS hardware) using the PCS simulation capabilities from initiation to termination.
These tests include both inference inputs and intermediate VC-matrix-computation variables, allowing direct comparison between the suite's Jacobians and VC matrices and those produced from the TensorFlow emulators. The latter are used within the FreeGSNKE Pulse Design Tool's (FPDT) \cite{pentland_freegsnke_2026} closed-loop simulation environment, which reproduces the PCS Category framework in Python.  

We selected MASTU shots $53000$ (high elongation) and $53010$ (super-X divertor). 
Shot playback generated the RTVC-predicted shape parameters and the VC matrices that would have been computed had RTVC Algorithm been configured for these shots during operation. 
Table \ref{tab:test-scenarios} summarises the different scenarios developed to assess the robustness of the deployed PCS implementation. 
A baseline configuration was established for each plasma shot, with subsequent tests introducing controlled variations in the inference time window, plasma current density profile parameters, inverse matrix configuration, and shape parameter override selection. 
This approach enabled each configurable aspect of the implementation to be validated independently while maintaining reproducible test conditions.

\begin{center}
\begin{minipage}{\textwidth}
\centering
\label{tab:test-scenarios}

\small
\setlength{\tabcolsep}{4pt}
\renewcommand{\arraystretch}{1.15}

\begin{tabular}{
    p{0.45cm}
    p{0.85cm}
    p{1.55cm}
    p{3.30cm}
    p{4.30cm}
    p{4.10cm}
}
\toprule
\textbf{ID} &
\textbf{Shot} &
\textbf{Interval (s)} &
\textbf{Inverse configuration} &
\textbf{Override configuration} &
\textbf{Purpose} \\
\midrule

1 &
53000 &
0.2--0.6 &
All seven emulated shape parameters &
All controlled parameters: Shape1--Shape4 &
Baseline configuration for shot 53000. \\

2* &
53000 &
0.2--0.6 &
All seven emulated shape parameters &
All controlled parameters &
Use of an alternative profile definition. \\

3 &
53000 &
0.2--0.6 &
Squareness gap, nose gap and strike point excluded &
All controlled parameters &
Variation of the parameters included in the pseudo-inverse. \\

4 &
53000 &
0.2--0.6 &
All seven emulated shape parameters &
Shape1 (\(R_{\mathrm{out}}\)) not overridden &
Single-parameter override variation. \\

5 &
53000 &
0.2--0.6 &
All seven emulated shape parameters &
Shape1 (\(R_{\mathrm{out}}\)) and Shape3 (\(Z_x\)) not overridden &
Multiple-parameter override variation. \\

6 &
53010 &
0.2--0.8 &
All seven emulated shape parameters &
Shape1--Shape4 and Div-T5 &
Baseline configuration for shot 53010. \\

7 &
53010 &
0.2--0.8 &
Squareness gap and nose gap excluded &
Strike-point parameters Div1, Div3 and Div4 not overridden &
Combined inverse and override variation. \\

\bottomrule
\end{tabular}
\captionsetup{labelfont=bf}
\captionof{table}{Summary of software validation scenarios used during simulation testing. *Scenario 2 was ran using plasma current density profiles from shot 53010.}
\end{minipage}
\end{center}

\begin{figure}
    \centering
    \includegraphics[height=0.96\textheight]{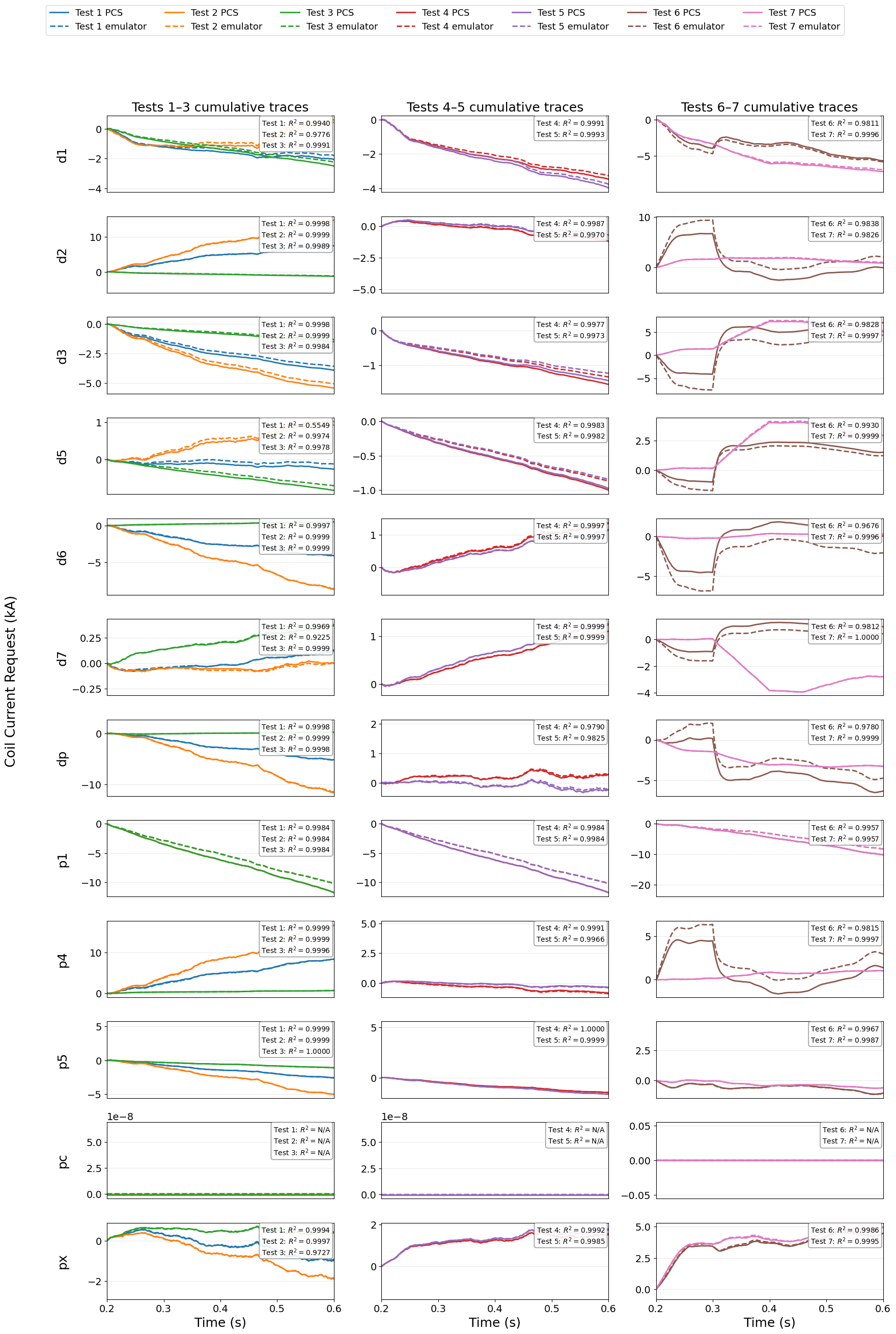}
    \caption{Scenarios 1-7 of PCS Simulation testing, comparing the coil current requests produced by TensorFlow emulators in FPDT (``emulator'') against the requests produced by RTVC (``PCS'').}
    \label{fig:test1-9}
\end{figure}

Figure \ref{fig:test1-9} displays the results from each of these tests, focusing on the cumulative integral of the PF coil current requests. Results of the shot playback (with the RTVC Algorithm enabled) are compared to those of the TensorFlow emulators embedded with the FPDT. 
We consistently observe $R^2 \geq 0.95$ in the majority of tests, with any offsets attributable to the PCS simulation environment rather than to the controller itself. 
Within PCS, each input channel has individual analogue-to-digital converter calibration parameters (offset and scaling), whereas simulation replay applies a single global offset and scaling factor to all channels. 
This calibration mismatch accumulates through the control calculations, producing an offset most visible in the final coil current requests, as seen in the plots.
Alongside this testing, we also measured inference performance (i.e. how often the real-time VCs were updated throughout the shot), achieving a bounded cycle time of $5$--$5.6$ms.

\section{MAST-U Commissioning Shots}
\label{s-irl-testing}

Before deploying the RTVC Algorithm for active plasma shape control on MAST-U, we ran it in parallel (i.e. ``piggyback'' mode) to the conventional VC Algorithm on real MAST-U shots.
Using the production-ready PCS hardware, the VC Algorithm produced the real-time coil current derivative requests used to control the plasma, while the RTVC Algorithm logged requests it would have produced to control the plasma, given the real-time input plasma measurements.
These requests could then be directly compared to those issued by the TensorFlow VC emulators.

\begin{figure}
    \centering
    \includegraphics[width=0.715\linewidth]{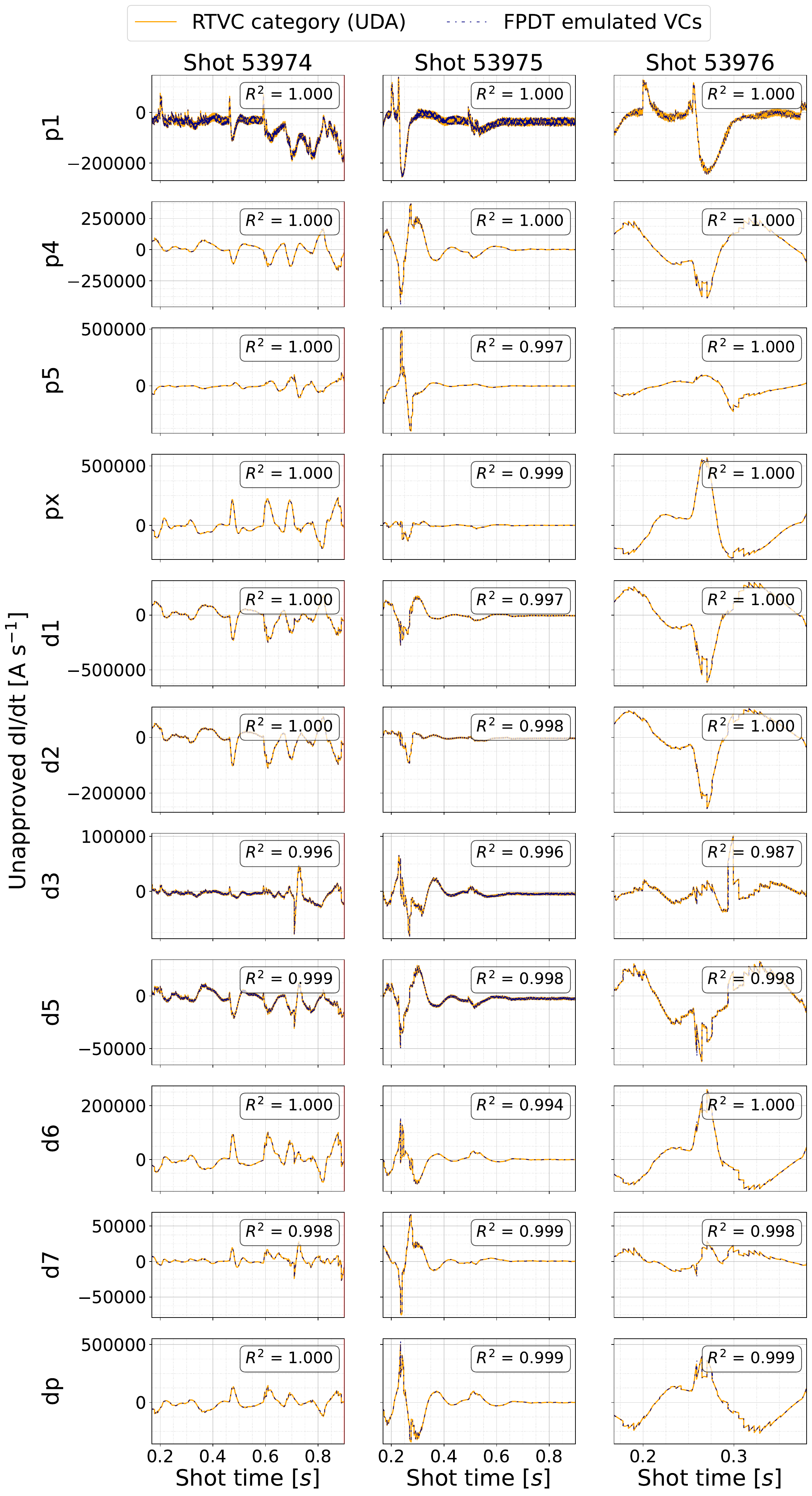}
    \caption{Unapproved dI/dt requests in commissioning shots 53974, 53975 and 53976 from Python emulators and RTVC. Note that shot 53976 terminated at 0.38s---unrelated to RTVC.}
    \label{fig:unapproved-didt-traces}
\end{figure}

This commissioning was performed over three different shots: 53974, 53975, and 53976. 
The ``unapproved'' (i.e. prior to machine safety limit clipping) coil current derivative requests from both RTVC and the FPDT can be found in Figure \ref{fig:unapproved-didt-traces}. 
The $R^2$ values show almost perfect agreement between the two sets of requests, completing the validation and integration of RTVC within the MAST-U PCS. 

As with the integration testing, the VC update latency during the commissioning shots was confirmed to lie within $[5,5.6]$ms, corresponding to approximately $130$-$144$ real-time, state-dependent VCs being generated per shot (for full $1$s shot cases). 

\section{Conclusions}
\label{s-conclusion}

This paper outlines how emulators trained to predict a tokamak plasma shape
(given plasma currents, coil currents, and plasma profile parameters) were
translated into a real-time control scheme and integrated into the MAST-U PCS.
We demonstrate that this real-time scheme issues the same commands as the
TensorFlow-based emulators in a closed-loop control workflow, establishing
readiness for plasma control via the RTVC Algorithm.

We have also characterised the real-time execution latency of the RTVC
workflow. The measured timings indicate that the present implementation
is compatible with experimental testing on MAST-U under the conditions
considered here, while leaving margin for further optimisation of
inference and communication overheads.

RTVC has been used successfully to actively control MAST-U plasmas across a
range of scenarios, as described separately \cite{Amorisco2026}. These results inform
future development of the RTVC Server and Algorithm, including further safety
constraints and improved handling of unexpected deviations in control.

By demonstrating real-time appropriate latencies, with an inference server capable of running general neural network models, this work lays the groundwork for other full-precision models to be used in control functions on tokamaks with system-on-a-chip PCS environments, where an accuracy-latency tradeoff (which may be required on accelerated hardware) is currently not necessary.

\section{Acknowledgements}

This work was funded by the Fusion Computing Lab collaboration (between UKAEA and STFC Hartree) and part funded by the EPSRC Energy Programme (EP/W006839/1). 

\printbibliography

\clearpage

\end{document}